\documentclass[
aps,%
12pt,%
final,%
notitlepage,%
oneside,%
onecolumn,%
nobibnotes,%
nofootinbib,%
noshowpacs,%
centertags]%
{revtex4}
\graphicspath{{figs/}}

\begin{document}

\noindent {\it Astronomy Reports, 2013, ü 5}\bigskip \hrule  \vspace{15mm}

\title{\uppercase{The Luminosity function of Narrow-Line Seyfert galaxies based on SDSS data}}

\author{\firstname{\bf \copyright $\:$ 2013 \quad  A.~A.}~\surname{Ermash}}
\affiliation{Astro Space Centre, P.N. Lebedev Physical Institute of Russian Academy of Sciences,\\ Moscow, Russia}

\begin{abstract}
\medskip\centerline{\footnotesize Received  xx.xx.2012;$\;$ Accepted xx.xx.2012}\bigskip\bigskip\noindent
We present measurements of AGN type 1 luminosity function in the forbidden line [OIII]5007\AA~using data from SDSS DR7. 
A special attention is paid to a subclass of Seyfert galaxies~--- the Narrow-Line Seyfert galaxies. These galaxies have relatively narrow broad permitted emission lines with FWHM $\leq 2000$ km/s.  
A new approach in calculation of the luminosity function is presented. 
We also account for the large-scale structure variations of the Universe density. 
The results obtained are compared with the ones from the literature. 
A prediction of X-ray luminosity function based on our results shows an agreement with observations. 
One of our preliminary conclusions is that NLSy1 seem to occupy a more narrow range in the nuclear luminosity than BLSy1, but the average values are within errors.

\end{abstract}

\maketitle

\vspace{34mm}  \hrule  \vspace{4mm}
\noindent {\it\footnotesize E-mail:}  \qquad
\url{aermash@gmail.com}     

\newpage

\section{Introduction}

It is not yet completely clear how do active nuclei (AGN) and their host galaxies evolve. 
During the recent years there is a growing tendency to pay a close attention to secular processes. Such processes are quite slow and are caused by different types of internal instabilities in contrast to mergings. 
For example, \cite{letawe07,schawinski10} showed that there is a high fraction of systems with disk-like morphology of host galaxies among high-redshift quasars and with a high amount of gas in them. 

According to \cite{dave11}, at redshifts $z\sim2$, many of the galaxies with ``clumpy'' morphology that thought to be the results of mergings show regular velocity maps similar to the normal disk galaxies.
The observed morphology is a result of a very active star formation caused by gas infall along the large-scale filaments. 

It is believed that the growth of a central SuperMassive Black Hole (SMBH) must be connected with the growth of the host galaxy or with some of its components~--- bulge, halo etc.
Many researchers reported a very tight correlation between mass of the spherical components of galaxies and masses of their black holes (see, for example, the classical paper \cite{magorrian98})
It is also accepted that there is a correlation between the mass of central black hole and the velocity dispersion of stars in bulge: $M_{BH}\propto\sigma_\ast^n$, where $n>1$ (see, for example, \cite{gebhardt00,ferrarese00}).
A ``Spherical component'' is the whole galaxy for elliptical systems and the bulge for disk galaxies.

Nevertheless, many galaxies have the so-called ``pseudobulges''.
They have lower $B/T$ ratios compared to classical bulges, low Sersic indexes ($n_b<2$) and also high angular momenta. 
Such bulges are dynamically more similar to disks of galaxies than to spherical non-rotating systems.
It is believed that pseudobulges are formed via secular evolution, which in turn can be triggered by different internal processes or by outer influences such as flyby of a satellite galaxy or a minor merging \cite{elichemoral06}. 
It is very important to stress that these bulges do not follow the well-known correlations $M_{BH}$ -- $M_{bulge}$ and $M_{BH}$ -- $\sigma_\ast$.

Narrow-Line Seyfert galaxies are good representatives of such type of systems. They were first recognized as a distinct type of active galaxies in \cite{osterbrock85} as Seyfert galaxies having unusually narrow broad permitted lines in their spectra ($FWHM\leq2000$ km/s).
Ordinary Seyfert galaxies in context of this classification are called Broad-Line Seyfert galaxies (BLS).
NLS show many other peculiar properties. 
Host galaxies of NLS are, on average, of later types than those of BLS. The average Hubble Types of host galaxies are $\langle T\rangle=3.0$ and $\langle T\rangle=1.0$ for NLS and BLS, respectively \cite{deo06}.
Numerical encoding of the Hubble types is according to the Third Reference Catalogue (RC3) \cite{devaucouleurs91}. $T=1$ corresponds to Sa galaxies, $T=2$~--- Sab, $T=3$~--- Sb and so forth. 

Their host galaxies always have pseudobulges judging on NLS galaxies with available photometry.
On the other hand, the host galaxies of BLS may have classical bulges as well \cite{xivry11}.

It is now widely accepted that NLS accrete at high Eddington ratios, close to unity and that their black holes are undermassive when compared with those of BLS \cite{sobolewska11,xukomossa11}.
Properties of NLS in $\gamma$-, x-ray and radio deserve a special attention.
Many researchers have dedicated their studies to this problem. 
The main result of these studies is the recognition of the presence of relativistic jets in NLS.
The very existence of jets in late-type spiral galaxies with low mass black holes is an intriguing discovery.
Some authors supposed that high angular momentum of supermassive black holes is a factor that leads to the launch of relativistic jets in such systems.
Such a rapid rotation is acquired by a black hole via accretion of gas with a high angular momentum, transported to the galactic center by secular processes.

\section{Data Processing} 

For our study we use data from the SDSS DR 7 (Sloan Digital Sky Survey Data Release 7) \cite{abazajian09}. 
It needs to be mentioned that in our study we do not use diagnostic diagrams \cite{kewley01,kauffmann03} because there is no necessity in it.
This work is dedicated to AGN type I with widths of the broad permitted lines FWHM $\geq1200$ km/s.
Special methods of distinguishing there objects from SF (Star Forming) and TO (Transition Objects) are not required.
 
\begin{figure}
\setcaptionmargin{5mm}
\onelinecaptionsfalse
\centering
\begin{minipage}[b]{.49\linewidth}
\centering\includegraphics[width=\linewidth]{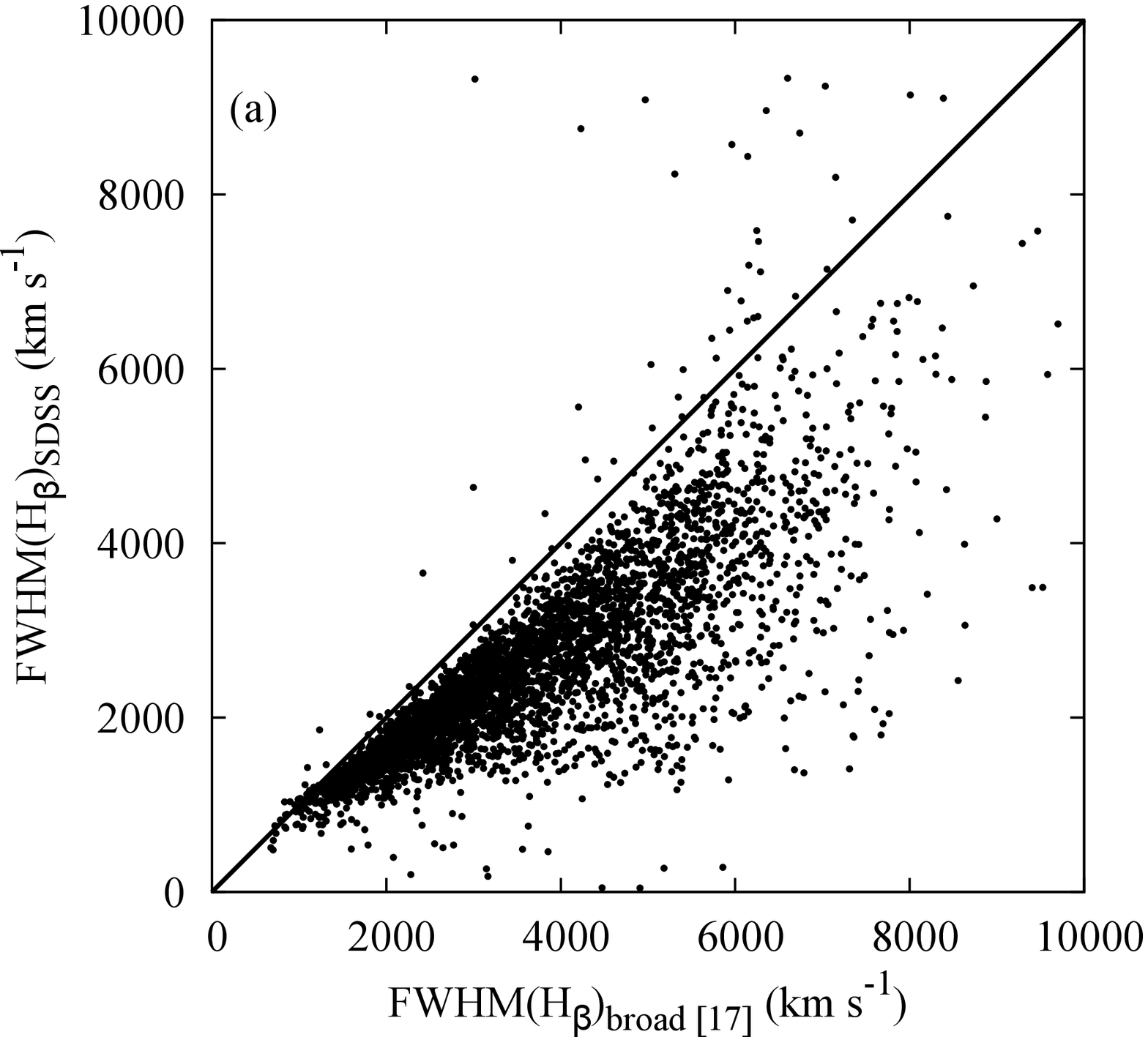}
\end{minipage}
\hfill
\begin{minipage}[b]{.49\linewidth}
\centering\includegraphics[width=\linewidth]{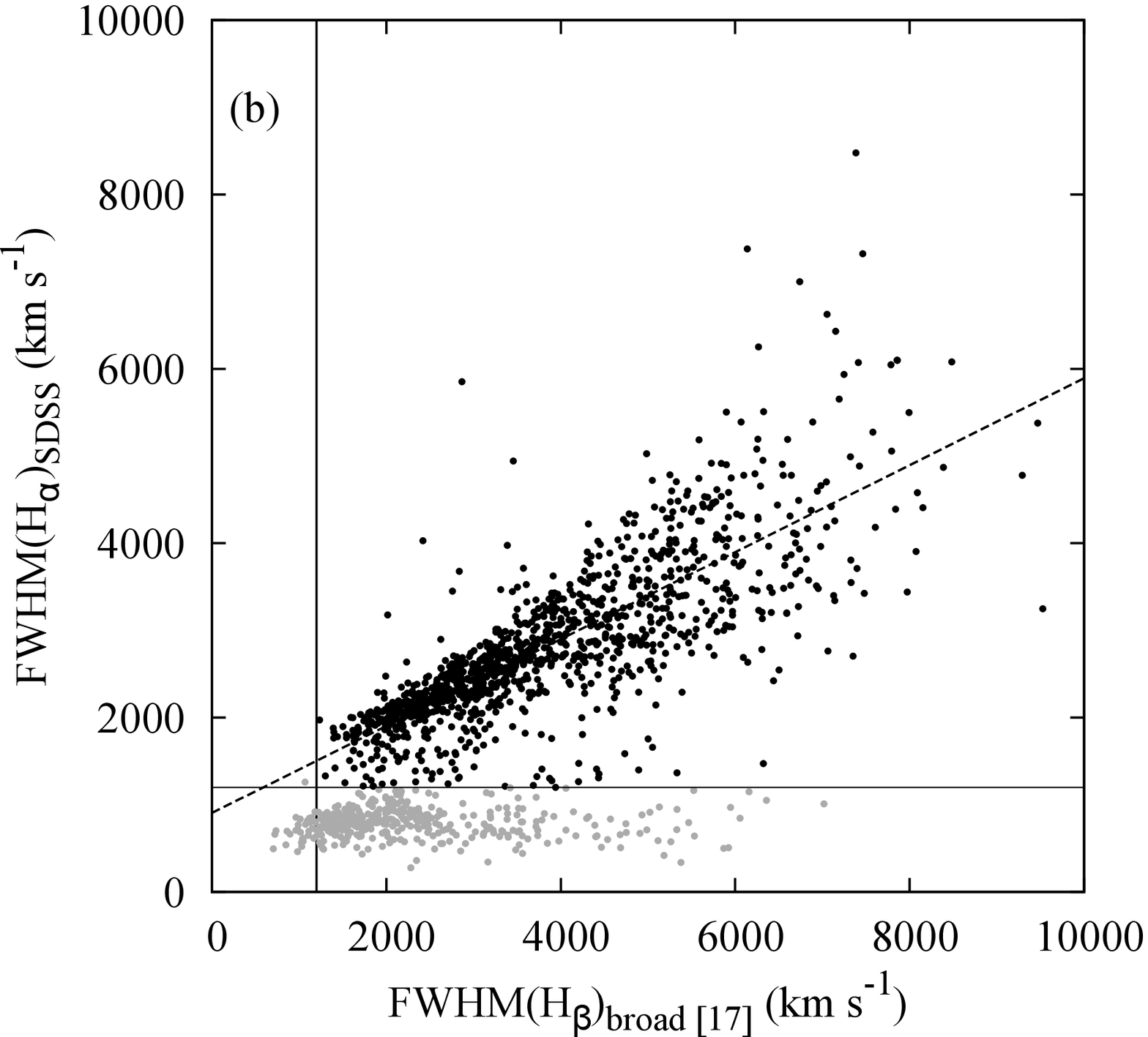}
\end{minipage}
\captionstyle{centerlast}
\begin{flushleft}
\caption
{
The dependence of the FWHM of the H$\beta$(a) and H$\alpha$(b) based on SDSS single Gaussian fitting on the FWHM of the H$\beta$ from \cite{hu08}. Grey dots indicate the objects without BLR for one of the lines, i.e.
${\rm FWHM}({\rm H}\beta)<1200$ km/s or 
${\rm FWHM}({\rm H}\alpha) <1200$ km/s (horizontal line).
\label{fig_fwhm}
} 
\end{flushleft}
\end{figure}

We use H$\alpha$ line instead of H$\beta$ to classify AGN into NLS and BLS.
It is known that H$\beta$ line has a prominent narrow component.
This is why it is impossible to use width estimations of this line based on a single Gaussian fit from SDSS catalogue to classify AGN.
In figure~\ref{fig_fwhm}a the ${\rm FWHM}({\rm H}\beta)_{\rm SDSS}$ is plotted versus ${\rm FWHM}({\rm H}\beta)_{broad}$.
${\rm FWHM}({\rm H}\beta)_{\rm SDSS}$ 
is the H$\beta$ width from SDSS survey,  
${\rm FWHM}({\rm H}\beta)_{broad}$
is the FWHM of the broad component of H$\beta$ line from \cite{hu08}, where authors thoroughly fitted the spectra. 
As can be seen from this figure, it is impossible to use widths of H$\beta$ line from SDSS.

Let us now direct our attention to the H$\alpha$ line.
It does not show such a prominent narrow component.
In figure~\ref{fig_fwhm}b 
${\rm FWHM}({\rm H}\alpha)_{\rm SDSS}$ is plotted versus
${\rm FWHM}({\rm H}\beta)_{broad}$.
It is necessary to exclude objects which have at least one line narrower than the threshold value $1200$~km/s. Objects without Broad Line Region are AGN type 2 and are not subject of this study.
As can be seen in fig.~\ref{fig_fwhm}b there is a tight correlation
${\rm FWHM}({\rm H}\alpha)_{\rm SDSS}=b\cdot{\rm FWHM}({\rm H}\beta)_{broad}+a$
with the following parameters:
\[a=0.499\pm0.011, \qquad b=906\pm46\,.\]
The deviation of $a$ coefficient from unity can be easily explained.
The excitation energies of H$\alpha$ and H$\beta$ lines are different.
That is why H$\alpha$ line will be emitted at larger distances from the active nucleus and will have lesser FWHM. 
For this study the crucial moment is the presence of a correlation which leads to a possibility to classify AGN into BL AGN (Broad-Line AGN) and NL AGN (Narrow-Line AGN) using the data on H$\alpha$ widths.  

The task of calculation of the intrinsic AGN luminosity is not among the ultimately solved ones.
All approaches described in literature can be divided in two groups: based on emission lines and continuum.
AGN luminosity can be estimated using continuum observations only when host galaxy contribution is negligible. For luminous quasars it is the case.
For example, such an approach was used in \cite{croom04}.
In order to measure luminosities of fainter AGN is it necessary to use photometric fitting.
In different papers were used different sets of components: Point Spread Function (PSF) + Sersic profile \cite{schawinski10}, PSF + two Sersic profiles with different $n_b$ \cite{mathur12}, PSF + de-Vaucouleur profile + exponential disk \cite{bennert11}.
But this approach has several significant drawbacks.
At first, many AGN lack photometric observations of sufficient quality to separate photometric components.
This is a significant problem for SDSS survey because resolution of it is not very high.
Secondly, proper accounting for orientation effects and internal absorption is a difficult and often an impossible task.

The use of emission lines to estimate AGN luminosity has some significant advantages.
In this study we utilize forbidden oxygen line [OIII]$\lambda$5007\AA.
In some papers it is stated that the contribution from star formation to this line is negligible (for example, \cite{haostrauss05}).
This line is formed in the Narrow-Line Region which is more distant from the nucleus than BLS and has a double-cone geometry. This implies that observed luminosity in [OIII] does not depend on orientation \cite{haostrauss05}.

\begin{figure}[t!]
\setcaptionmargin{5mm}
\onelinecaptionsfalse
\includegraphics[width=0.5\textwidth]{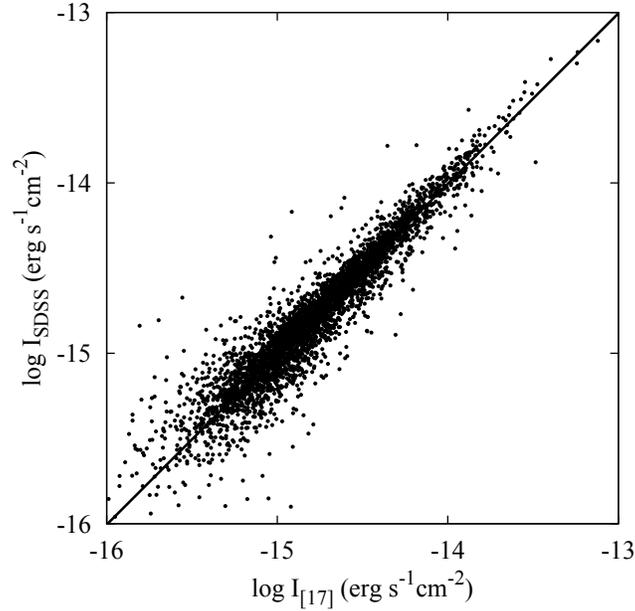}
\captionstyle{centerlast}
\caption
{
A comparison between integral intensity in [OIII] line obtained by the automatic pipeline of SDSS with the  intensities taken from \cite{hu08} obtained by a thorough spectral fitting.
\label{lhu_lew}
}
\end{figure}

In this study we do not perform the spectral fitting. It is necessary to check if it is possible to use line intensity data from SDSS.
In order to do so we plot $\lg I_{\rm [OIII],\,SDSS}$ versus $\lg I_{\rm [OIII],\,[17]}$ (see fig.~\ref{lhu_lew}).
$\lg I_{\rm [OIII],\,SDSS}$ is an [OIII] intensity from SDSS survey,
$\lg I_{\rm [OIII],\,[17]}$ is an [OIII] intensity obtained in \cite{hu08} by accurate spectral fits. 
The relation is as follows: $\lg I_{\rm [OIII],\,SDSS}=\lg I_{\rm [OIII],\,[17]}+a$
, where $a=-0.0053\pm0.0027$.
Due to the strength of the correlation [OIII] intensities obtained by the automatic SDSS pipeline can be used to calculate AGN luminosities.

We select total 9020 BL AGN objects from SDSS DR7 with  detectable [OIII] and H$\alpha$ emission having ${\rm FWHM(H\alpha)}\geq1200$ km/s.
We divide this sample into two subsamples, NLS and BLS.
There are 2082 and 6938 objects in these subsamples, respectively.
We set upper limit on the redshift equal to 0.18 because on higher redshifts it is impossible to apply our normalization algorithm (see details below).
The lower limit on the distance was set to $55$ Mpc to avoid the influence of the Local Supercluster.
In the redshift interval considered the spectral lines used in this study are within the interval of spectral sensitivity of the SDSS survey.

In this study we used the following cosmological parameters in accordance with SDSS DR7:
$\Omega_M=0.279$, $\Omega_L=0.721$, $h=0.701$.
Results of other studies were converted to this cosmology.

\section{The Luminosity Function}

The are two main methods of obtaining the luminosity function.
These are the maximum likelihood method and the $\dfrac{V}{V_{max}}$ method.
In the recent years due to substantial growth of the volumes of the catalogues there is a growing tendency to use a more simple, but having significant advantages $\dfrac{V}{V_{max}}$ method.
The main advantage is that this method is free from a priori presumptions about the shape of the luminosity function. 
The $\dfrac{V}{V_{max}}$ method implies calculation for each object the maximum volume where it can be detected by the observer. 
The task of calculating the $V_{max}$ from emission line observations is quite complex. 
Because there are several reasons why an object cannot be observed: visual magnitude is of limits of completeness of the survey, the decrease of $S/N$ with distance, widening of the lines at larger redshifts, pixel overfill in spectra of close luminous objects etc.
In order to apply the $\dfrac{V}{V_{max}}$ method one must create model spectra for each object, simulate the noise and then calculate maximum volume from which the object under consideration can be observed.
Another drawback is that it does not take into account variations of the Large Scale structure.
 
In order to use our method it is necessary to obtain a function of average probability of object detection $\rho(d_c)$, where $d_c$ is the comoving density. 
Hence, we can derive average value of $v_{max}$ for objects in the considered luminosity bin
\[\langle v_{max}\rangle=\int_{d_{c,\,min}}^{d_{c,\,max}}4\pi r^2\rho(d_c)dr\,,\]
where $d_{c,\,min}$ and $d_{c,\,max}$ are the lower and upper distance limits, respectively. 
The value of the luminosity function in the considered bin is:
\[\hat\phi(L)=\frac{N_{norm}}{\langle v_{max}\rangle\, d\lg{L_{\rm [OIII]}}}\,,\]
where $N_{norm}$ is the amount of galaxies after normalization.
The $\hat\phi(L)$ is expressed in the units of $Mpc^{-3}(\lg{L_{\rm [OIII]})^{-1}}$ i.e. the amount of objects per unit of volume per unit of the logarithm of the luminosity. 

The total error of $\hat\phi$ estimation consists of the Poisson error $\sigma_{N_{obs},poiss}=\sqrt{N_{obs}}$ and the error in estimating $\langle v_{max} \rangle$ ($\sigma_{\langle v_{max} \rangle}$) and is equal to:
\[\hat\phi(L)=\frac{1}{d \lg{L_{\rm [OIII]}}}\sqrt{\left(\frac{\theta \sigma_{N_{obs},poiss}}{\langle v_{max}\rangle }\right)^2+\left(\frac{N_{norm}\sigma_{\langle v_{max}\rangle}}{\langle v_{max}\rangle^2}\right)^2}\,,\]
where $\theta$ is a factor corresponding to the normalization, so $N_{norm}=\theta N_{obs}$.
In this study all the luminosities are expressed in units of  $L/L_\odot$, where $L_\odot=3.844\times10^{33}$~erg/s.
As is said above, the existing methods do not take into account the variations of the Large-Scale structure of the Universe.
In order to account for this effect we use the following approach.
 
The luminosity function of inactive galaxies for the Local Universe is well known and was obtained in many independent studies.
Let us define $\langle\rho_{gal}\rangle$ as the average density of galaxies calculated from the luminosity function and $\rho_{gal}$ as an observed density of galaxies in the considered volume.
The amount of AGN satisfying some predefined conditions is $N_{AGN}$.
This allows us to obtain the amount of AGN in the considered volume normalized to the average density of galaxies in the Local Universe:
\[N_{AGN,\,norm}=N_{AGN}\frac{\langle\rho_{gal}\rangle}{\rho_{gal}}\,.\]
In order to proceed the normalization let us create a sample of galaxies limited in magnitude by $m_{min}=14.5$ and $m_{max}=17.6$.
Within these limits the SDSS survey is believed to be complete.
Let us define $L_{obs}$ as observed total luminosity of inactive galaxies in the considered volume.
If $L_i$ is a luminosity of i-th galaxy, then $L_{obs}=\sum_{i=1}^{N}L_i$.
 
The luminosity function is usually fitted by
\[\phi(L)dL=\phi^*(L/L^*)^\alpha \exp{(-L/L^*)}d(L/L^*)\]
or in the absolute magnitudes \cite{schechter76}
\[\phi(M)dM=(0.4\ln 10)\phi^*10^{0.4(M^*-M)(1+\alpha)} \exp{\left(-10^{0.4(M^*-M)}\right)}dM\,.\]
In \cite{monterodorta09} the following parameters of the local luminosity function in $r$ band of the SDSS are obtained: $\phi^*=0.0090\pm0.0007$, $M^*-5\lg h=-20.73\pm0.04$, $\alpha=-1.23\pm0.02$
Using these parameters it is possible to obtain the calculated total luminosity of galaxies per unit of volume:
\[L_{sch}=\int_{L_1}^{L_2}L'\phi(L')dL\,.\]
Integration in this formula should have upper and lower limits $L_{max}$ and $L_{min}$ or $M_{max}$ and $M_{min}$.
It is because for faint galaxies the luminosity function is not well defined and when high value of $L_{max}$ is chosen the result of integration drastically depends on the selected parameters of the Schechter function.
In this study we used $M_{max}=-17$, $M_{min}=-23$ in the $r$ band. We also use these limits when we calculate the $L_{obs}$.
 
It is now possible to obtain the relation of the average density of galaxies in Local Universe to observed density of galaxies in considered volume:
\[\frac{\langle\rho_{gal}\rangle}{\rho_{gal}}=\frac{L_{sch} V \mu}{L_{obs}}=\kappa\,,\]
where V is the volume of the Universe within distance limits $d_{c,\,min}$,$d_{c,\,max}$ and $\mu$ is the coverage of the survey.
 
\begin{figure}[t!]
\setcaptionmargin{5mm}
\onelinecaptionsfalse
\centering
\begin{minipage}[b]{.50\linewidth}
\centering\includegraphics[width=\linewidth]{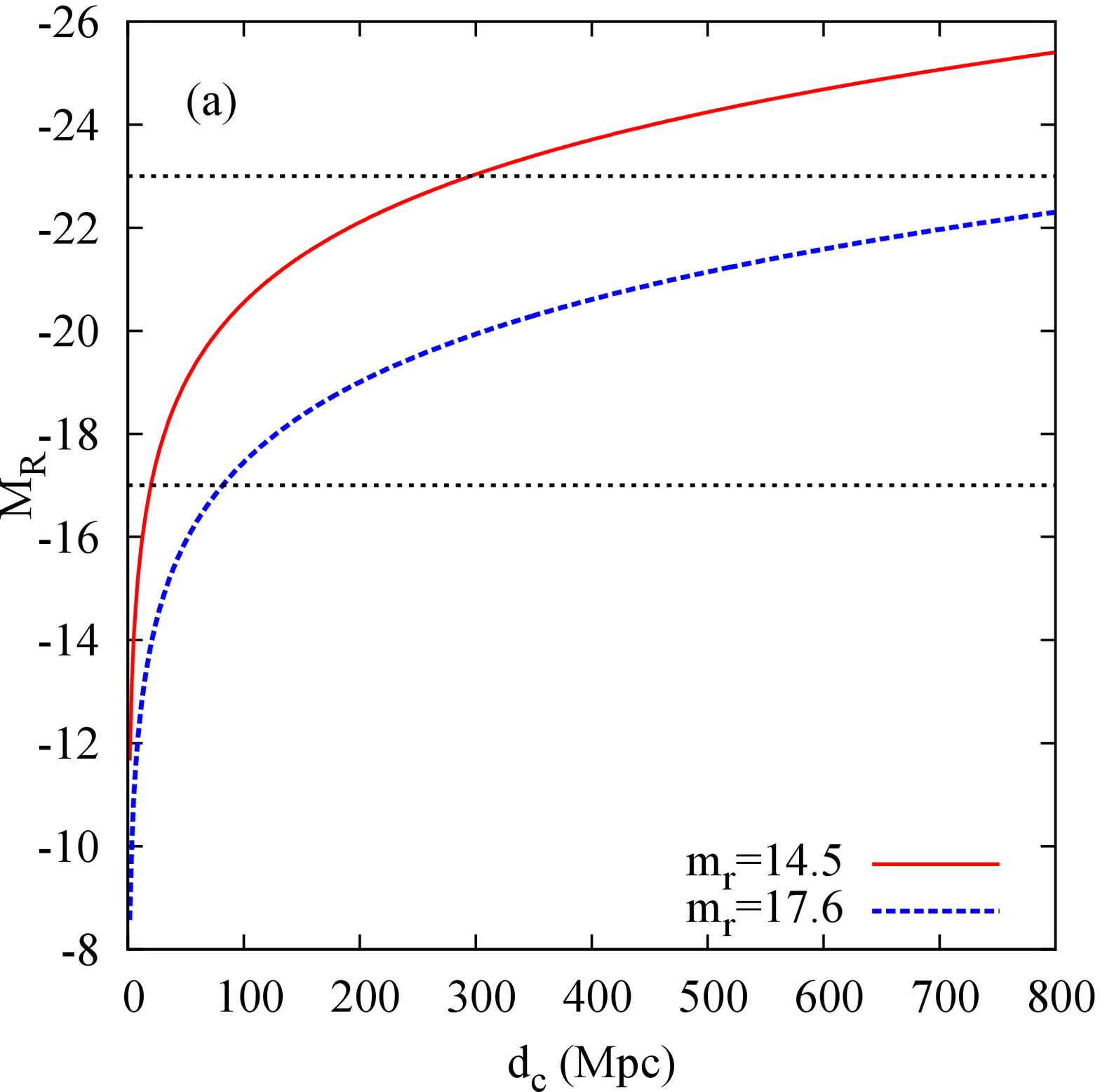}
\end{minipage}
\hfill
\begin{minipage}[b]{.48\linewidth}
\centering\includegraphics[width=\linewidth]{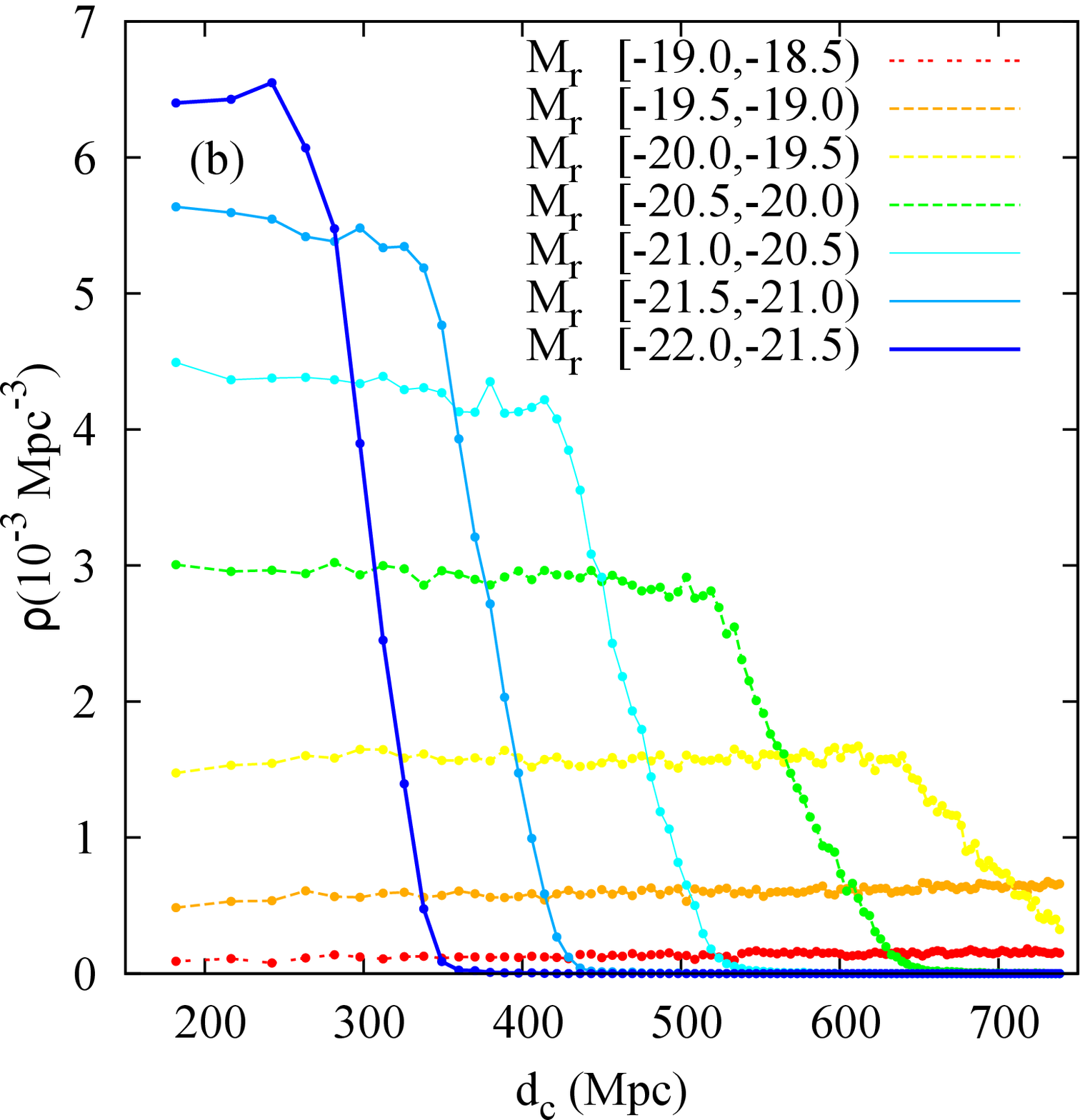}
\end{minipage}
\captionstyle{centerlast}
\caption
{
(a)~--- The dependence of absolute magnitudes on the comoving distances for visible magnitudes corresponding to the completeness limits of SDSS.
The horizontal dashed lines are the limiting absolute magnitudes used in this study for normalization.
See details in the text.
(b)~--- The dependence of normalized volume density on the comoving distance for different narrow intervals in the absolute magnitudes.
\label{limits_flat}
}
\end{figure}

Of course, this relation, is correct only when parameters of the Schechter function are constant. For galaxies in the Local Universe it is the case.
We consider the normalization possible if $L_{sch}/L_{sch,\,tot}$ is larger than threshold value $0.1$, where $L_{sch,\,tot}$ is the total luminosity of all galaxies, derived from Schechter function.
It quite interesting that when this approach is used it is not required to know the coverage area of the survey. Since when concentrations are used we have:
 \[n_{AGN,\,norm}=N_{AGN}\frac{L_{sch} V   \mu}{L_{obs}} \,\frac{1}{\mu V}=
N_{AGN}\frac{L_{sch} }{L_{obs}}\,.
\]
In means that $L_{obs}$ already contains the survey coverage in it.
Let us now check the reliability of the described algorithm.
In order to do so we apply it to a sample of galaxies with known absolute magnitudes.
As we have already mentioned above, the SDSS survey is complete in the magnitude interval $14.5<m_r<17.6$ \cite{abazajian09}.
The dependence of galaxy concentration on $z$ in $M_1<M<M_2$ interval can be expressed in the following form: $\rho(z)=\rho_*f_{gs}(z)f_c(z)$, where $f_{gs}(z)$ is a term that reflects variations of the Large Scale structure and $f_c$ is the completeness of the survey in selected interval on certain $z$.
After the normalization we obtain: $\rho(z)=\rho_*f_c(z)$.
If the normalization algorithm is working properly, then a plateau followed by a rapid decline when visual magnitude reaches the completeness limit(see fig.~\ref{limits_flat}a) of SDSS survey should be seen. This is the exact description of what can be seen in fig.~\ref{limits_flat}b.

The luminosity function of inactive galaxies can be obtained directly. The concentration of galaxies in absolute magnitude interval $M_1<M<M_2$ is equal to the plateau of the $\rho(z)$ function.
The situation with AGN is far more complex. The are many reasons why an object will not be in the AGN sample.
These reasons can be divided in two groups.
The first group consists of factors that make observation of close objects impossible (too bright visual magnitudes, very bright emission lines that leads to overfill of pixels in their spectra etc.).
The probability function $\rho_{inc}(d_c)$ will increase with distance.
Let us search it in the form $\exp\left(-b_1/d_c^2\right)$, where $b_1$ is a free parameter.
The second group consists of factors that cause the decrease of detection probability with distance due to decrease of the observed flux.
The corresponding function $\rho_{dec}$ will decrease with distance to the object.
Let us search it in the from of $\exp\left(-b_2 d_c^2\right)$, where $b_2$ is a free parameter.
 
Now we will briefly address the issue of misclassification.
The characteristic that defines the NLS and BLS populations is the width of broad permitted lines, in our case H$\alpha$. 
As was noted in \cite{denney09} when the signal to noise ratio is low FWHM of emission lines is systematically underestimated.
That is why there will be a tendency to misclassify a genuine BLS due to the underestimation of FWHM with increasing redshift.
Some of the genuine NLS will be misclassified because their measured FWHM will actually be below the threshold between Sy1 and Sy2.
Also some of objects which are actually BLS will be classified as NLS at higher redshifts.
All these effects will contribute to considered probability functions.
 
For each luminosity bin $L_{\rm [OIII]}\in[L_i,L_{i+1}]$ the normalized AGN density must be fitted with the function $\rho_{AGN}=a\rho_{dec}\rho_{inc}$.
The corresponding average value of $v_{max}$ is: 
\[\langle v_{max}\rangle=\int_{d_{c,min}}^{d_{c,max}}4\pi\ d_c^2 e^{-b/d_c^2} e^{-c d_c^2} d(d_c)\]
 
Because the observed values of $\rho_{AGN}(z)$ are actually small, especially for luminous objects we use cumulative function:
\[f_{obs}(d_{c})=N_{AGN}(d_{c,i}<d_{c})\,,\]
where $N_{AGN}(d_{c,\,i}<d_{c})$ is an amount of AGN in the considered luminosity bin within distance limits $d_{c,min}<d_{c,\,i}<d_{c}$.
For fitting we use the following function:
\[f_{fit}(d_{c,\,i})=\int_{d_{c,\,min}}^{d_{c,\,i}}4\pi a d_c^2 e^{\left(-\frac{b_1}{d_c^2}\right)} e^{(-b_2 d_c^2)}d(d_c)\,.\]
 
It should be noted that the main result of the approximation is the shape of the function $\rho_{AGN}(d_c)$.
The value $\sigma_{\langle v_{max}\rangle}$ is derived from the error of the approximation, i.e. from $\sigma(a)$, $\sigma(b_1)$ and $\sigma(b_2)$ (see below).

We use neither a fixed step in luminosity nor in the distance.
The bin widths were selected manually because the luminosity function varies drastically with $L_{\rm [OIII]}$.
Bin widths and amounts of objects in every bin are listed in table \ref{tab_lf}.
A fixed step in distance makes it impossible to use our normalization algorithm at low distances.
The volume in each bin depends on the distance and the amount of objects in close bins is actually low.
Instead of it we use a fixed step in volume $\delta V_c$.
For the $i$-th bin we have the distance limits:
 \[\sqrt[3]{\frac{4}{3}\pi d_{c,0}^3+i\delta V_c}~\frac{3}{4}\pi<d_c<\sqrt[3]{\frac{4}{3}\pi d_{c,0}^3+(i+1)\delta V_c}~\frac{3}{4}\pi\,.\]
Finally, we obtain luminosity functions for three samples: all Sy1, NLSy1, BLSy1.
 
All calculations are done using programs written in PYTHON language by the author of this paper. 
 
\section{A comparison of the obtained Luminosity function with ones from the literature} 
\begin{figure}[t!]
\setcaptionmargin{5mm}
\onelinecaptionsfalse
\includegraphics[width=0.9\textwidth]{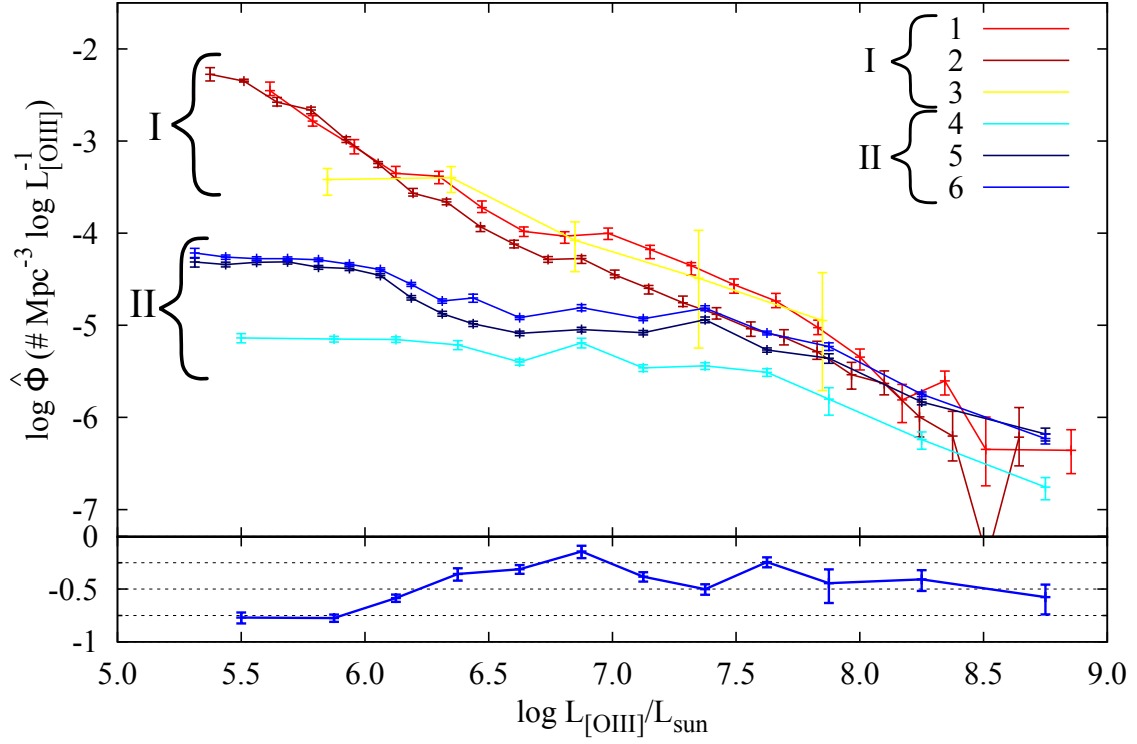}
\captionstyle{centerlast}
\caption
{
The luminosity functions of NLS, BLS and Sy1(NLS + BLS) obtained in this study.
For the sake of comparison LFs from \cite{haostrauss05} and \cite{bongiorno10} obtained in the same line [OIII] $\lambda5007 \mbox{\AA}$ are also plotted:
{\it 1}~--- Sy1 \cite{haostrauss05}.
{\it 2}~--- Sy2 \cite{haostrauss05}.
{\it 3}~--- AGN type 2 \cite{bongiorno10}.
{\it 4, 5, 6}~--- NLS, BLS, Sy1(NLS + BLS), respectively, obtained in this study.
Group I consists of luminosity functions from the literature. Group II~--- luminosity functions obtained in this study.
\label{fig_lf_oiii}
}
\end{figure}

The luminosity functions of NLSy1, BLSy1 and all Sy1 are shown in fig.~\ref{fig_lf_oiii}.
The results obtained are also listed in table~\ref{tab_lf}.
The luminosity function from \cite{haostrauss05} and \cite{bongiorno10} are showed for comparison.
In \cite{haostrauss05} LFs were obtained for Sy1 and Sy2, in \cite{bongiorno10} only for Sy2.
As can be seen in fig.~\ref{fig_lf_oiii}, the luminosity functions from \cite{haostrauss05,bongiorno10} are in a good agreement with each other.
Our luminosity function is significantly lower.

But, the luminosity functions from many different studies show disagreements with \cite{haostrauss05}.
In order to compare our results with as many as possible LFs calculated by different authors we convert our LF from [OIII]$\lambda$5007\AA~to the Johnson band B according to \cite{bongiorno10}:
\[\lg{\left(\frac{L_{\rm [OIII]}}{L_\odot}\right)}=-0.38M_B-0.42\,,\]
In order to convert luminosities from H$\alpha$ to $M_B$ we use the relation from \cite{schulze09}:
\[M_B=-2.1(\lg{L_{\rm H\alpha}}-42)-20.1\,.\]

It should be noted that these relations have significant dispersions.
For example, it results in disagreement between luminosity functions converted to $B$ band from H$\alpha$ and [OIII] from the same study \cite{haostrauss05}.

\begin{figure}[t!]
\setcaptionmargin{5mm}
\onelinecaptionsfalse
\includegraphics[width=0.9\textwidth]{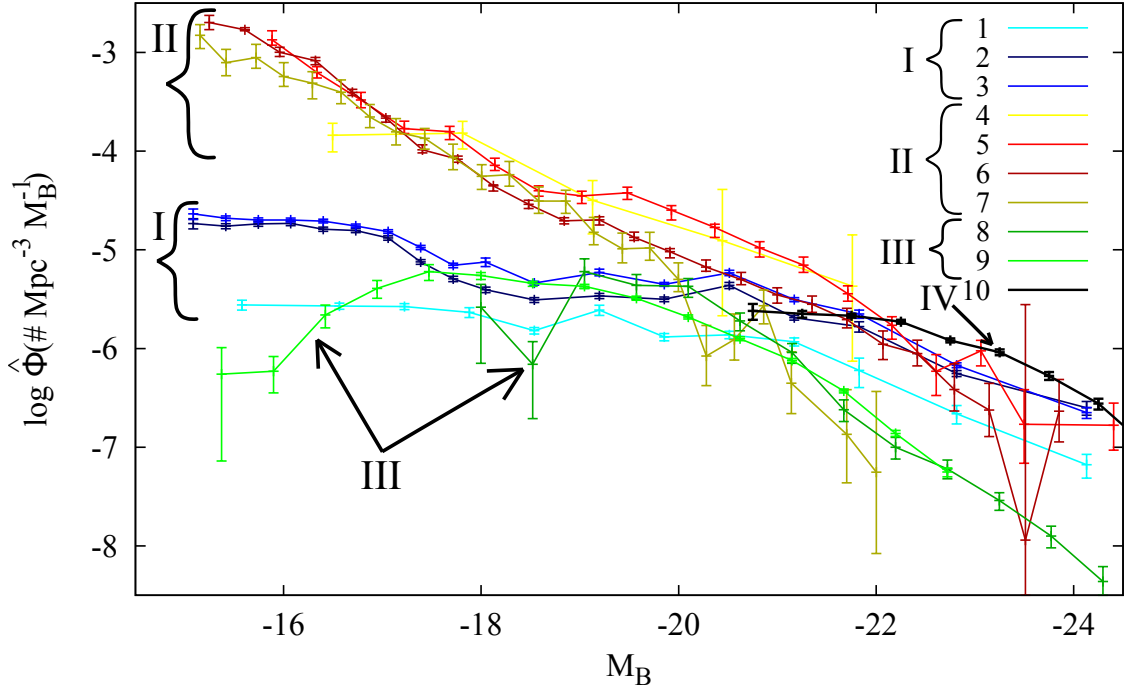}
\captionstyle{centerlast}
\caption
{
The luminosity functions in the $B$ band obtained by converting luminosities from [OIII] $\lambda5007 \mbox{\AA}$ or H$\alpha$:
\textit {1, 2, 3} -- NLSy1, BLSy1, NLSy1 + BLSy1, respectively, obtained in this study.
\textit {4} -- AGN type 2 from [OIII] \cite{bongiorno10}.
\textit {5} -- Sy1 from [OIII] \cite{haostrauss05}.
\textit {6} -- Sy2 from [OIII] \cite{haostrauss05}.  
\textit {7} -- Sy1 + Sy2 from H$\alpha$ \cite{haostrauss05}.
\textit {8} -- Sy1 from H$\alpha$ \cite{schulze09}.
\textit {9} -- Sy1 from H$\alpha$ \cite{greeneho09}.
\textit {10} -- QSO, obtained in $B$ band \cite{croom04}.
The luminosity functions are divided in four groups. Group I consists of LFs obtained in this study.
Group II~--- LFs from \cite{haostrauss05,bongiorno10}. Group III~--- LFs from \cite{schulze09,greeneho07}. LFs in groups II and III show a good agreement with each other. Group IV~--- the LF of quasars \cite{croom04}.
\label{fig_lf_mb}
}
\end{figure}

The result of conversion of luminosity functions is shown in fig.~\ref{fig_lf_mb}, where the following LFs are also plotted:
\begin{enumerate}
 \item Converted from [OIII] to $M_B$ (\cite{bongiorno10,haostrauss05} and this study)
 \item Converted from $H_\alpha$ \cite{haostrauss05,schulze09,greeneho07}
 \item The quasar luminosity function at redshift $0.4<z<0.88$ measured in B band \cite{croom04}.
\end{enumerate} 
There is a known erratum in \cite{greeneho07}, we used corrected data published by authors in \cite{greeneho09}.

As can be seen from these figs LFs obtained by different authors form two distinct groups (groups II and III in fig.~\ref{fig_lf_mb}).
They consist of the results of \cite{haostrauss05,bongiorno10} and \cite{schulze09,greeneho07}, respectfully.
LF of Sy1 obtained by us is between these contradicting results.
It is important that our LF agrees well with the LF of quasars \cite{croom04} on highest luminosities.

It is remarkable that $\phi(NLS)/\phi(BLS)$ is not constant.
Instead, it is actually a function of the luminosity (fig.~\ref{fig_lf_oiii}., bottom panel).
It has a maximum between $\log{(L_{[OIII]}/L_\odot)}=6.8$ and $7.6$.
At larger luminosities it slightly decreases.
The decrease towards lower luminosities is $\sim0.5dex$ to $\log{(L_{OIII}/L_\odot)}=5.5$.

A putative explanation can be proposed when the results of \cite{xukomossa12} are considered.
The authors studied samples of NLS and BLS selected from SDSS DR3.
The AGN luminosities were estimated using $M_{r,\,{\rm PSF}}$ and $M_{g,\,{\rm PSF}}$.
That is why only objects with $L_{AGN}\gg L_{HG}$ were included in the sample.
The ones of the key results of their work are the distributions of $\log{(L/L_{edd})}$ and $\log{(M_{BH}/M_\odot)}$.
As was stated in \cite{xukomossa12} and many other studies, the distributions of $M_{BH}$ and $L/L_{edd}$ differ significantly for NLS and BLS.

Due to the fact that $\lg{L}=\lg{\left(L_{edd}\,\,\dfrac{L}{L_{edd}}\right)}=
\lg{\left(1.3\times10^{38}\right)}+\lg{\dfrac{M_{BH}}{M_\odot}}+\lg{\dfrac{L}{L_{edd}}}$, if $\lg{\dfrac{M_{BH}}{M_\odot}}$ and $\lg{\dfrac{L}{L_{edd}}}$ have normal distributions $\lg{L}$ also has a normal distribution.

The parameters of $\lg{L_{bol}}$ distributions are:\\
for NLS:
\[\mu=44.589\pm0.052,\]
\[\sigma=0.404\pm0.043,\]
for BLSy1:
\[\mu=44.535\pm0.088,\]
\[\sigma=0.560\pm0.061,\]
where $\mu$ and $\sigma$ are the expected value and dispersion.
The expected values are within the errors, but dispersions differ significantly.
Thus, $\phi(NLS)/\phi(BLS)$ ratio has a maximum at $\log{(L_{bol})}=44.64$ which corresponds to $\log{(L_{[OIII]}/L_\odot)}=7.4$.
The dispersion of NLS distribution is lower, therefore NLS occupy a narrower range in luminosity than BLS.

We would like to stress it again that the comparison with \cite{xukomossa12} is putative. 
In \cite{xukomossa12} $L_{AGN}$ was estimated using $M_{r,\,{\rm SDSS}}$ and $M_{g,\,{\rm SDSS}}$, i.e. the distribution of $L_{AGN}$ in their sample is different from this study.

\begin{figure}[t!]
\setcaptionmargin{5mm}
\onelinecaptionsfalse
\includegraphics[width=0.9\textwidth]{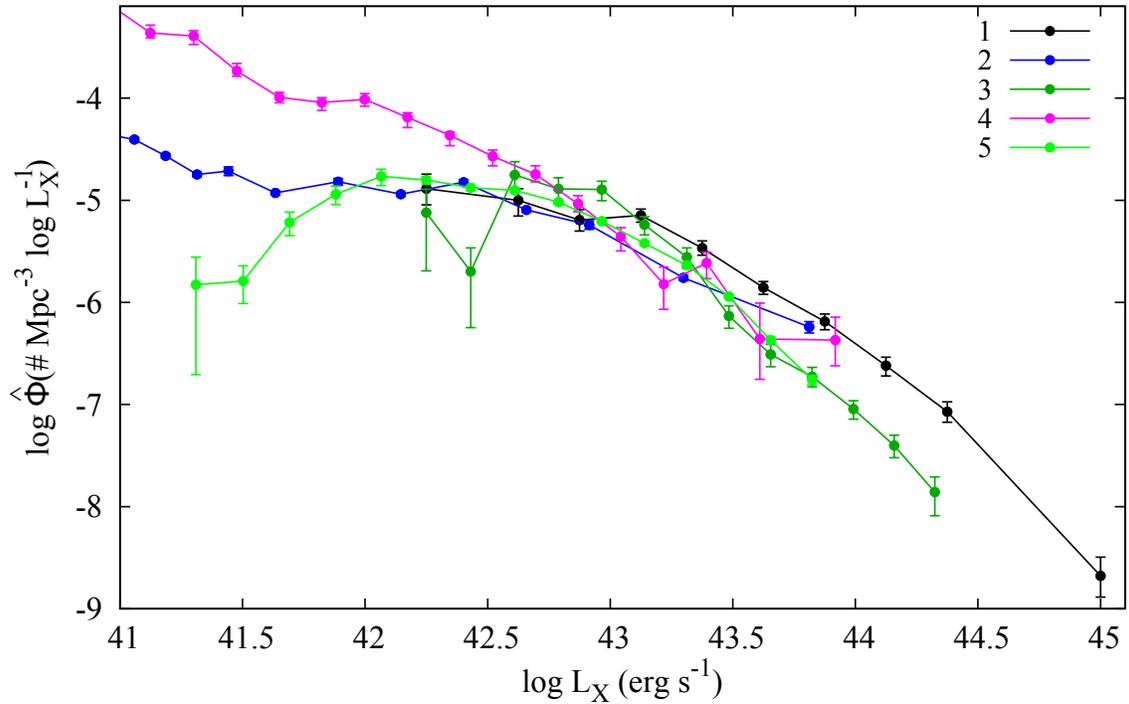}
\captionstyle{centerlast}
\caption
{
The observed Sy1 x-ray Luminosity Function (0.5--2 kev) \cite{hasinger05}.
The predicted Sy1 x-ray LFs based on the luminosity functions from this study and the ones from the literature are also plotted:
\textit 1 -- x-ray 0.5--2 keV \cite{hasinger05}.
\textit 2 -- LF prediction based on [OIII] LF from this study.
\textit 3 -- [OIII] \cite{haostrauss05}.
\textit 4 -- H$\alpha$ \cite{schulze09}.
\textit 5 -- H$\alpha$ \cite{greeneho09}.
\label{fig_lf_xray}
}

\end{figure}

We calculate a predicted soft x-ray(0.5--2kev) luminosity function based on the $L_{[OIII]}$ data.
The luminosities in [OIII]$\lambda$5007\AA~are converted into the soft x-ray following \cite{marconi04,heckman05}.
The result is plotted in fig.~\ref{fig_lf_xray}.
The x-ray LF of Sy1 based on the ROSAT, XMM-Newton and CHANDRA data \cite{hasinger05} is also shown in fig.~\ref{fig_lf_xray}.
For a comparison the predictions based on the $H_\alpha$ luminosity functions from \cite{schulze09,greeneho09} and [OIII] from \cite{haostrauss05} are also plotted in fig.~\ref{fig_lf_xray}.
$H_\alpha$ luminosities are converted to the bolometric ones following \cite{greeneho07}, then from the bolometric ones to $L_{0.5-2kev}$ following \cite{marconi04}.
 
It should be stressed that our prediction corresponds to the actually observed x-ray luminosity function better than the LFs from other studies. 

\section{Conclusions} 

\begin{itemize}
\item We developed a method of evaluation of AGN luminosity function based on emission-line data accounting for variations of the density of galaxies due to the large-scale structure. The [OIII]$\lambda$5007\AA~emission-line luminosity functions for NLS, BLS, Sy1 (NLS+BLS) are obtained.
\item In order to compare our results with luminosity functions from other studies we converted our LF to the $B$ band. LF of Sy1 obtained by us is between contradicting results from different publications.
\item At largest luminosities the LF of Sy1 agrees very well with the one of quasars \cite{croom04}.
\item A prediction of soft x-ray luminosity function based on our emission-line LF is in an agreement with the observed x-ray LF.
\item $\phi(NLS)/\phi(BLS)$ is not a constant, but a function of luminosity.
It has a maximum between $\log{(L_{[OIII]}/L_\odot)}=6.8$ and $7.6$.
This agrees with the results of \cite{xukomossa12}.
The average luminosities of NLS and BLS are equal within errors, but widths of the distributions differ.
NLS occupy a narrower range of the AGN luminosity.

\end{itemize}

\medskip
The author would like to acknowledge B.V.Komberg and O.V.Verkhodanov for interesting discussions and useful comments.

This work was supported by the Basic Research Program of the Presidium of the Russian Academy of Sciences ``Nonstationary Processes in Objects in the Universe''(P-21) and
by The Ministry of education and science of Russian Federation, project 8405
and The Program of state support of Leading Scientific Schools in Russian Federation (grant SC-2915.2012.2 ``The formation of the Large-Scale structure of the Universe and cosmological processes'').

Funding for the SDSS and SDSS-II has been provided by the Alfred P. Sloan Foundation, 
the Participating Institutions, the National Science Foundation, the U.S. 
Department of Energy, the National Aeronautics and Space Administration, 
the Japanese Monbukagakusho, the Max Planck Society, and the Higher Education 
Funding Council for England. The SDSS Web Site is http://www.sdss.org/.

The SDSS is managed by the Astrophysical Research Consortium for 
the Participating Institutions. The Participating Institutions are 
the American Museum of Natural History, Astrophysical Institute Potsdam, 
University of Basel, University of Cambridge, Case Western Reserve University, 
University of Chicago, Drexel University, Fermilab, the Institute for Advanced 
Study, the Japan Participation Group, Johns Hopkins University, the Joint Institute 
for Nuclear Astrophysics, the Kavli Institute for Particle Astrophysics and Cosmology, 
the Korean Scientist Group, the Chinese Academy of Sciences (LAMOST), 
Los Alamos National Laboratory, the Max-Planck-Institute for Astronomy (MPIA), 
the Max-Planck-Institute for Astrophysics (MPA), New Mexico State University, 
Ohio State University, University of Pittsburgh, University of Portsmouth, 
Princeton University, the United States Naval Observatory, and the University of Washington.

This research has made use of CosmoPY package for PYTHON http://roban.github.com/CosmoloPy/. 

\bibliography{paper_lf}

\begin{thebibliography}{32}
\expandafter\ifx\csname natexlab\endcsname\relax\def\natexlab#1{#1}\fi
\expandafter\ifx\csname bibnamefont\endcsname\relax
  \def\bibnamefont#1{#1}\fi
\expandafter\ifx\csname bibfnamefont\endcsname\relax
  \def\bibfnamefont#1{#1}\fi
\expandafter\ifx\csname citenamefont\endcsname\relax
  \def\citenamefont#1{#1}\fi
\expandafter\ifx\csname url\endcsname\relax
  \def\url#1{\texttt{#1}}\fi
\expandafter\ifx\csname urlprefix\endcsname\relax\def\urlprefix{URL }\fi
\providecommand{\bibinfo}[2]{#2}
\providecommand{\eprint}[2][]{\url{#2}}

\bibitem[{\citenamefont{Letawe \emph{et~al.}}(2007)\citenamefont{Letawe,
  Magain, Courbin \emph{et~al.}}}]{letawe07}

\refitem{article}
\bibinfo{author}{\bibfnamefont{G.}~\bibnamefont{Letawe}},
  \bibinfo{author}{\bibfnamefont{P.}~\bibnamefont{Magain}},
  \bibinfo{author}{\bibfnamefont{F.}~\bibnamefont{Courbin}},
  \bibnamefont{\emph{et~al.}}, \bibinfo{journal}{Monthly Not. Roy. Astron.
  Soc.} \textbf{\bibinfo{volume}{378}}, \bibinfo{pages}{83}
  (\bibinfo{year}{2007}).

\bibitem[{\citenamefont{Schawinski
  \emph{et~al.}}(2010)\citenamefont{Schawinski, Treister, Urry
  \emph{et~al.}}}]{schawinski10}

\refitem{article}
\bibinfo{author}{\bibfnamefont{K.}~\bibnamefont{Schawinski}},
  \bibinfo{author}{\bibfnamefont{E.}~\bibnamefont{Treister}},
  \bibinfo{author}{\bibfnamefont{C.~M.} \bibnamefont{Urry}},
  \bibnamefont{\emph{et~al.}}, \bibinfo{journal}{Astrophys. J. (Letters)}
  \textbf{\bibinfo{volume}{727}}, \bibinfo{pages}{L31} (\bibinfo{year}{2010}).

\bibitem[{\citenamefont{Dave}(2011)}]{dave11}

\refitem{article}
\bibinfo{author}{\bibfnamefont{R.}~\bibnamefont{Dave}},
  \bibinfo{journal}{e-Print arXiv:1101.5397v1 [astro-ph]}
  (\bibinfo{year}{2011}).

\bibitem[{\citenamefont{Magorrian \emph{et~al.}}(1998)\citenamefont{Magorrian,
  Tremaine, Richstone \emph{et~al.}}}]{magorrian98}

\refitem{article}
\bibinfo{author}{\bibfnamefont{J.}~\bibnamefont{Magorrian}},
  \bibinfo{author}{\bibfnamefont{S.}~\bibnamefont{Tremaine}},
  \bibinfo{author}{\bibfnamefont{D.}~\bibnamefont{Richstone}},
  \bibnamefont{\emph{et~al.}}, \bibinfo{journal}{Astron. J.}
  \textbf{\bibinfo{volume}{115}}, \bibinfo{pages}{2285} (\bibinfo{year}{1998}).

\bibitem[{\citenamefont{Gebhardt \emph{et~al.}}(2000)\citenamefont{Gebhardt,
  Bender, Bower \emph{et~al.}}}]{gebhardt00}

\refitem{article}
\bibinfo{author}{\bibfnamefont{K.}~\bibnamefont{Gebhardt}},
  \bibinfo{author}{\bibfnamefont{R.}~\bibnamefont{Bender}},
  \bibinfo{author}{\bibfnamefont{G.}~\bibnamefont{Bower}},
  \bibnamefont{\emph{et~al.}}, \bibinfo{journal}{Astrophys. J. (Letters)}
  \textbf{\bibinfo{volume}{539}}, \bibinfo{pages}{L13} (\bibinfo{year}{2000}).

\bibitem[{\citenamefont{Ferrarese and Merritt}(2000)}]{ferrarese00}

\refitem{article}
\bibinfo{author}{\bibfnamefont{L.}~\bibnamefont{Ferrarese}} \bibnamefont{and}
  \bibinfo{author}{\bibfnamefont{D.}~\bibnamefont{Merritt}},
  \bibinfo{journal}{Astrophys. J. (Letters)} \textbf{\bibinfo{volume}{539}},
  \bibinfo{pages}{L9} (\bibinfo{year}{2000}).

\bibitem[{\citenamefont{Eliche-Moral
  \emph{et~al.}}(2006)\citenamefont{Eliche-Moral, Balcells, Aguerri, and
  Gonz{\'a}lez-Garc{\'i}a}}]{elichemoral06}

\refitem{article}
\bibinfo{author}{\bibfnamefont{M.~C.} \bibnamefont{Eliche-Moral}},
  \bibinfo{author}{\bibfnamefont{M.}~\bibnamefont{Balcells}},
  \bibinfo{author}{\bibfnamefont{J.~A.~L.} \bibnamefont{Aguerri}},
  \bibnamefont{and} \bibinfo{author}{\bibfnamefont{A.~C.}
  \bibnamefont{Gonz{\'a}lez-Garc{\'i}a}}, \bibinfo{journal}{Astron. Astrophys.}
  \textbf{\bibinfo{volume}{457}}, \bibinfo{pages}{91} (\bibinfo{year}{2006}).

\bibitem[{\citenamefont{Osterbrock and Pogge}(1985)}]{osterbrock85}

\refitem{article}
\bibinfo{author}{\bibfnamefont{D.~E.} \bibnamefont{Osterbrock}}
  \bibnamefont{and} \bibinfo{author}{\bibfnamefont{R.~W.} \bibnamefont{Pogge}},
  \bibinfo{journal}{Astrophys. J.} \textbf{\bibinfo{volume}{297}},
  \bibinfo{pages}{166} (\bibinfo{year}{1985}).

\bibitem[{\citenamefont{Deo \emph{et~al.}}(2006)\citenamefont{Deo, Crenshaw,
  and Kraemer}}]{deo06}

\refitem{article}
\bibinfo{author}{\bibfnamefont{R.~P.} \bibnamefont{Deo}},
  \bibinfo{author}{\bibfnamefont{D.~M.} \bibnamefont{Crenshaw}},
  \bibnamefont{and} \bibinfo{author}{\bibfnamefont{S.}~\bibnamefont{Kraemer}},
  \bibinfo{journal}{Astron. J.} \textbf{\bibinfo{volume}{132}},
  \bibinfo{pages}{321} (\bibinfo{year}{2006}).

\bibitem[{\citenamefont{de~Vaucouleurs
  \emph{et~al.}}(1991)\citenamefont{de~Vaucouleurs, de~Vaucouleurs, Corwin
  \emph{et~al.}}}]{devaucouleurs91}

\refitem{book}
\bibinfo{author}{\bibfnamefont{G.}~\bibnamefont{de~Vaucouleurs}},
  \bibinfo{author}{\bibfnamefont{A.}~\bibnamefont{de~Vaucouleurs}},
  \bibinfo{author}{\bibfnamefont{H.~G.~J.} \bibnamefont{Corwin}},
  \bibnamefont{\emph{et~al.}}, \emph{\bibinfo{title}{Third Reference Catalogue
  of Bright Galaxies}} (\bibinfo{publisher}{New York: Springer},
  \bibinfo{year}{1991}).

\bibitem[{\citenamefont{{Orban de Xivry}
  \emph{et~al.}}(2011)\citenamefont{{Orban de Xivry}, Davies, Schartmann
  \emph{et~al.}}}]{xivry11}

\refitem{article}
\bibinfo{author}{\bibfnamefont{G.}~\bibnamefont{{Orban de Xivry}}},
  \bibinfo{author}{\bibfnamefont{R.}~\bibnamefont{Davies}},
  \bibinfo{author}{\bibfnamefont{M.}~\bibnamefont{Schartmann}},
  \bibnamefont{\emph{et~al.}}, \bibinfo{journal}{Monthly Not. Roy. Astron.
  Soc.} \textbf{\bibinfo{volume}{417}}, \bibinfo{pages}{2721}
  (\bibinfo{year}{2011}).

\bibitem[{\citenamefont{Sobolewska
  \emph{et~al.}}(2011)\citenamefont{Sobolewska, Siemiginowska, and
  Gierlinski}}]{sobolewska11}

\refitem{article}
\bibinfo{author}{\bibfnamefont{M.~A.} \bibnamefont{Sobolewska}},
  \bibinfo{author}{\bibfnamefont{A.}~\bibnamefont{Siemiginowska}},
  \bibnamefont{and}
  \bibinfo{author}{\bibfnamefont{M.}~\bibnamefont{Gierlinski}},
  \bibinfo{journal}{Monthly Not. Roy. Astron. Soc.}
  \textbf{\bibinfo{volume}{413}}, \bibinfo{pages}{2259} (\bibinfo{year}{2011}).

\bibitem[{\citenamefont{Xu and Komossa}(2011)}]{xukomossa11}

\refitem{inproceedings}
\bibinfo{author}{\bibfnamefont{D.}~\bibnamefont{Xu}} \bibnamefont{and}
  \bibinfo{author}{\bibfnamefont{S.}~\bibnamefont{Komossa}}, in
  \emph{\bibinfo{booktitle}{Proceedings of the Workshop ``Narrow-Line Seyfert 1
  Galaxies and Their Place in the Universe'', PoS(NLS1) 006}}, edited by
  \bibinfo{editor}{\bibfnamefont{L.}~\bibnamefont{Foschini}},
  \bibinfo{editor}{\bibfnamefont{M.}~\bibnamefont{Colpi}},
  \bibinfo{editor}{\bibfnamefont{L.}~\bibnamefont{Gallo}},
  \bibnamefont{\emph{et~al.}} (\bibinfo{publisher}{Trieste, Italy: Proceedings
  of Science}, \bibinfo{year}{2011}).

\bibitem[{\citenamefont{Abazajian \emph{et~al.}}(2009)\citenamefont{Abazajian,
  Adelman-McCarthy, Ag{\"u}eros' \emph{et~al.}}}]{abazajian09}

\refitem{article}
\bibinfo{author}{\bibfnamefont{K.~N.} \bibnamefont{Abazajian}},
  \bibinfo{author}{\bibfnamefont{J.~K.} \bibnamefont{Adelman-McCarthy}},
  \bibinfo{author}{\bibfnamefont{M.~A.} \bibnamefont{Ag{\"u}eros'}},
  \bibnamefont{\emph{et~al.}}, \bibinfo{journal}{Astrophys. J. Suppl. Ser.}
  \textbf{\bibinfo{volume}{182}}, \bibinfo{pages}{543} (\bibinfo{year}{2009}).

\bibitem[{\citenamefont{Kewley \emph{et~al.}}(2001)\citenamefont{Kewley,
  Dopita, Sutherland \emph{et~al.}}}]{kewley01}

\refitem{article}
\bibinfo{author}{\bibfnamefont{L.~J.} \bibnamefont{Kewley}},
  \bibinfo{author}{\bibfnamefont{M.~A.} \bibnamefont{Dopita}},
  \bibinfo{author}{\bibfnamefont{R.~S.} \bibnamefont{Sutherland}},
  \bibnamefont{\emph{et~al.}}, \bibinfo{journal}{Astrophys. J.}
  \textbf{\bibinfo{volume}{556}}, \bibinfo{pages}{121} (\bibinfo{year}{2001}).

\bibitem[{\citenamefont{Kauffmann \emph{et~al.}}(2003)\citenamefont{Kauffmann,
  Heckman, Tremonti \emph{et~al.}}}]{kauffmann03}

\refitem{article}
\bibinfo{author}{\bibfnamefont{G.}~\bibnamefont{Kauffmann}},
  \bibinfo{author}{\bibfnamefont{T.~M.} \bibnamefont{Heckman}},
  \bibinfo{author}{\bibfnamefont{C.}~\bibnamefont{Tremonti}},
  \bibnamefont{\emph{et~al.}}, \bibinfo{journal}{Monthly Not. Roy. Astron.
  Soc.} \textbf{\bibinfo{volume}{346}}, \bibinfo{pages}{1055}
  (\bibinfo{year}{2003}).

\bibitem[{\citenamefont{Hu \emph{et~al.}}(2008)\citenamefont{Hu, Wang, Ho
  \emph{et~al.}}}]{hu08}

\refitem{article}
\bibinfo{author}{\bibfnamefont{C.}~\bibnamefont{Hu}},
  \bibinfo{author}{\bibfnamefont{J.-M.} \bibnamefont{Wang}},
  \bibinfo{author}{\bibfnamefont{L.~C.} \bibnamefont{Ho}},
  \bibnamefont{\emph{et~al.}}, \bibinfo{journal}{Astrophys. J.}
  \textbf{\bibinfo{volume}{687}}, \bibinfo{pages}{78} (\bibinfo{year}{2008}).

\bibitem[{\citenamefont{Croom \emph{et~al.}}(2004)\citenamefont{Croom, Smith,
  Boyle \emph{et~al.}}}]{croom04}

\refitem{article}
\bibinfo{author}{\bibfnamefont{S.~M.} \bibnamefont{Croom}},
  \bibinfo{author}{\bibfnamefont{R.~J.} \bibnamefont{Smith}},
  \bibinfo{author}{\bibfnamefont{B.~J.} \bibnamefont{Boyle}},
  \bibnamefont{\emph{et~al.}}, \bibinfo{journal}{Monthly Not. Roy. Astron.
  Soc.} \textbf{\bibinfo{volume}{349}}, \bibinfo{pages}{1397}
  (\bibinfo{year}{2004}).

\bibitem[{\citenamefont{Mathur \emph{et~al.}}(2012)\citenamefont{Mathur,
  Fields, Peterson, and Grupe}}]{mathur12}

\refitem{article}
\bibinfo{author}{\bibfnamefont{S.}~\bibnamefont{Mathur}},
  \bibinfo{author}{\bibfnamefont{D.}~\bibnamefont{Fields}},
  \bibinfo{author}{\bibfnamefont{B.~M.} \bibnamefont{Peterson}},
  \bibnamefont{and} \bibinfo{author}{\bibfnamefont{D.}~\bibnamefont{Grupe}},
  \bibinfo{journal}{e-Print arXiv:1101.5397v1 [astro-ph]}
  (\bibinfo{year}{2012}).

\bibitem[{\citenamefont{Bennert \emph{et~al.}}(2011)\citenamefont{Bennert,
  Auger, Treu, Woo \emph{et~al.}}}]{bennert11}

\refitem{article}
\bibinfo{author}{\bibfnamefont{V.~N.} \bibnamefont{Bennert}},
  \bibinfo{author}{\bibfnamefont{M.~W.} \bibnamefont{Auger}},
  \bibinfo{author}{\bibfnamefont{T.}~\bibnamefont{Treu}},
  \bibinfo{author}{\bibfnamefont{J.-H.} \bibnamefont{Woo}},
  \bibnamefont{\emph{et~al.}}, \bibinfo{journal}{Astrophys. J.}
  \textbf{\bibinfo{volume}{742}}, \bibinfo{pages}{107} (\bibinfo{year}{2011}).

\bibitem[{\citenamefont{Hao \emph{et~al.}}(2005)\citenamefont{Hao, Strauss, Fan
  \emph{et~al.}}}]{haostrauss05}

\refitem{article}
\bibinfo{author}{\bibfnamefont{L.}~\bibnamefont{Hao}},
  \bibinfo{author}{\bibfnamefont{M.~A.} \bibnamefont{Strauss}},
  \bibinfo{author}{\bibfnamefont{X.}~\bibnamefont{Fan}},
  \bibnamefont{\emph{et~al.}}, \bibinfo{journal}{Astron. J.}
  \textbf{\bibinfo{volume}{129}}, \bibinfo{pages}{1795} (\bibinfo{year}{2005}).

\bibitem[{\citenamefont{Schechter}(1976)}]{schechter76}

\refitem{article}
\bibinfo{author}{\bibfnamefont{P.~L.} \bibnamefont{Schechter}},
  \bibinfo{journal}{Astrophys. J.} \textbf{\bibinfo{volume}{203}},
  \bibinfo{pages}{297} (\bibinfo{year}{1976}).

\bibitem[{\citenamefont{Montero-Dorta and Prada}(2009)}]{monterodorta09}

\refitem{article}
\bibinfo{author}{\bibfnamefont{A.~D.} \bibnamefont{Montero-Dorta}}
  \bibnamefont{and} \bibinfo{author}{\bibfnamefont{F.}~\bibnamefont{Prada}},
  \bibinfo{journal}{Monthly Not. Roy. Astron. Soc.}
  \textbf{\bibinfo{volume}{399}}, \bibinfo{pages}{1106} (\bibinfo{year}{2009}).

\bibitem[{\citenamefont{Denney \emph{et~al.}}(2009)\citenamefont{Denney,
  Peterson, Dietrich \emph{et~al.}}}]{denney09}

\refitem{article}
\bibinfo{author}{\bibfnamefont{K.~D.} \bibnamefont{Denney}},
  \bibinfo{author}{\bibfnamefont{B.~M.} \bibnamefont{Peterson}},
  \bibinfo{author}{\bibfnamefont{M.}~\bibnamefont{Dietrich}},
  \bibnamefont{\emph{et~al.}}, \bibinfo{journal}{Astrophys. J.}
  \textbf{\bibinfo{volume}{692}}, \bibinfo{pages}{246} (\bibinfo{year}{2009}).

\bibitem[{\citenamefont{Bongiorno \emph{et~al.}}(2010)\citenamefont{Bongiorno,
  Mignoli, Zamorani \emph{et~al.}}}]{bongiorno10}

\refitem{article}
\bibinfo{author}{\bibfnamefont{A.}~\bibnamefont{Bongiorno}},
  \bibinfo{author}{\bibfnamefont{M.}~\bibnamefont{Mignoli}},
  \bibinfo{author}{\bibfnamefont{G.}~\bibnamefont{Zamorani}},
  \bibnamefont{\emph{et~al.}}, \bibinfo{journal}{Astron. Astrophys.}
  \textbf{\bibinfo{volume}{510}}, \bibinfo{pages}{56} (\bibinfo{year}{2010}).

\bibitem[{\citenamefont{Schulze \emph{et~al.}}(2009)\citenamefont{Schulze,
  Wisotzki, and Husemann}}]{schulze09}

\refitem{article}
\bibinfo{author}{\bibfnamefont{A.}~\bibnamefont{Schulze}},
  \bibinfo{author}{\bibfnamefont{L.}~\bibnamefont{Wisotzki}}, \bibnamefont{and}
  \bibinfo{author}{\bibfnamefont{B.}~\bibnamefont{Husemann}},
  \bibinfo{journal}{Astron. Astrophys.} \textbf{\bibinfo{volume}{507}},
  \bibinfo{pages}{781} (\bibinfo{year}{2009}).

\bibitem[{\citenamefont{Greene and Ho}(2007)}]{greeneho07}

\refitem{article}
\bibinfo{author}{\bibfnamefont{J.~E.} \bibnamefont{Greene}} \bibnamefont{and}
  \bibinfo{author}{\bibfnamefont{L.~C.} \bibnamefont{Ho}},
  \bibinfo{journal}{Astrophys. J.} \textbf{\bibinfo{volume}{667}},
  \bibinfo{pages}{131} (\bibinfo{year}{2007}).

\bibitem[{\citenamefont{Greene and Ho}(2009)}]{greeneho09}

\refitem{article}
\bibinfo{author}{\bibfnamefont{J.~E.} \bibnamefont{Greene}} \bibnamefont{and}
  \bibinfo{author}{\bibfnamefont{L.~C.} \bibnamefont{Ho}},
  \bibinfo{journal}{Astrophys. J.} \textbf{\bibinfo{volume}{704}},
  \bibinfo{pages}{1743} (\bibinfo{year}{2009}).

\bibitem[{\citenamefont{Xu \emph{et~al.}}(2012)\citenamefont{Xu, Komossa, Zhou
  \emph{et~al.}}}]{xukomossa12}

\refitem{article}
\bibinfo{author}{\bibfnamefont{D.}~\bibnamefont{Xu}},
  \bibinfo{author}{\bibfnamefont{S.}~\bibnamefont{Komossa}},
  \bibinfo{author}{\bibfnamefont{H.}~\bibnamefont{Zhou}},
  \bibnamefont{\emph{et~al.}}, \bibinfo{journal}{Astron. J.}
  \textbf{\bibinfo{volume}{143}}, \bibinfo{pages}{15} (\bibinfo{year}{2012}).

\bibitem[{\citenamefont{Marconi \emph{et~al.}}(2004)\citenamefont{Marconi,
  Risaliti, Gilli \emph{et~al.}}}]{marconi04}

\refitem{article}
\bibinfo{author}{\bibfnamefont{A.}~\bibnamefont{Marconi}},
  \bibinfo{author}{\bibfnamefont{G.}~\bibnamefont{Risaliti}},
  \bibinfo{author}{\bibfnamefont{R.}~\bibnamefont{Gilli}},
  \bibnamefont{\emph{et~al.}}, \bibinfo{journal}{Monthly Not. Roy. Astron.
  Soc.} \textbf{\bibinfo{volume}{351}}, \bibinfo{pages}{169}
  (\bibinfo{year}{2004}).

\bibitem[{\citenamefont{Heckman \emph{et~al.}}(2005)\citenamefont{Heckman,
  Ptak, Hornschemeier, and Kauffmann}}]{heckman05}

\refitem{article}
\bibinfo{author}{\bibfnamefont{T.~M.} \bibnamefont{Heckman}},
  \bibinfo{author}{\bibfnamefont{A.}~\bibnamefont{Ptak}},
  \bibinfo{author}{\bibfnamefont{A.}~\bibnamefont{Hornschemeier}},
  \bibnamefont{and}
  \bibinfo{author}{\bibfnamefont{G.}~\bibnamefont{Kauffmann}},
  \bibinfo{journal}{Astrophys. J.} \textbf{\bibinfo{volume}{634}},
  \bibinfo{pages}{161} (\bibinfo{year}{2005}).

\bibitem[{\citenamefont{Hasinger \emph{et~al.}}(2005)\citenamefont{Hasinger,
  Miyaji, and Schmidt}}]{hasinger05}

\refitem{article}
\bibinfo{author}{\bibfnamefont{G.}~\bibnamefont{Hasinger}},
  \bibinfo{author}{\bibfnamefont{T.}~\bibnamefont{Miyaji}}, \bibnamefont{and}
  \bibinfo{author}{\bibfnamefont{M.}~\bibnamefont{Schmidt}},
  \bibinfo{journal}{Astron. Astrophys.} \textbf{\bibinfo{volume}{441}},
  \bibinfo{pages}{417} (\bibinfo{year}{2005}).

\end{thebibliography}

\newpage

\begin{table*}
\setcaptionmargin{0mm} \onelinecaptionsfalse
\captionstyle{center}
\caption{The luminosity functions of NLS, BLS and Sy1(NLS + BLS). 
The units are as following. $L_{\rm [OIII]}$ is in units of $\lg{L/L_\odot}$,
Bin (size of the luminosity bun) is in dex,
luminosity function $\hat\phi(L)$ in $Mpc^{-3}(\lg{L_{\rm [OIII]})^{-1}}$,
$N$ is an amount of objects in every luminosity bin.
}
\label{tab_lf}

\bigskip
\small
\begin{tabular}{|c|c|c|c||c|c|c|c||c|c|c|c|}
\hline\noalign{\smallskip}
\multicolumn{4}{|c||}{NLSy1} &\multicolumn{4}{c||}{BLSy1} & \multicolumn{4}{c|}{Sy1} \\

\noalign{\smallskip}
\hline\noalign{\smallskip}
$L_{\rm [OIII]}$ &Bin &$\hat\phi(L)$ & $N$&
$L_{\rm [OIII]}$ &Bin &$\hat\phi(L)$ & $N$&
$L_{\rm [OIII]}$ &Bin &$\hat\phi(L)$ & $N$ \\

\noalign{\smallskip}
\hline\noalign{\smallskip}
5.5  &0.5   &$-5.138^{+0.047}_{-0.053}$  &256&5.3125&0.125 &$-4.313^{+0.048}_{-0.055}$ &299&5.3125&0.125  &$-4.215^{+0.050}_{-0.057}$ &339   \\
5.875&0.25  &$-5.150^{+0.030}_{-0.032}$  &228&5.4375&0.125 &$-4.340^{+0.024}_{-0.026}$ &323&5.4375&0.125  &$-4.260^{+0.023}_{-0.024}$ &374   \\
6.125&0.25  &$-5.154^{+0.029}_{-0.031}$  &230&5.5625&0.125 &$-4.317^{+0.023}_{-0.024}$ &388&5.5625&0.125  &$-4.277^{+0.021}_{-0.022}$ &452   \\
6.375&0.25  &$-5.213^{+0.047}_{-0.052}$  &211&5.6875&0.125 &$-4.312^{+0.022}_{-0.023}$ &439&5.6875&0.125  &$-4.277^{+0.020}_{-0.021}$ &524   \\
6.625&0.25  &$-5.399^{+0.031}_{-0.034}$  &193&5.8125&0.125 &$-4.370^{+0.020}_{-0.021}$ &469&5.8125&0.125  &$-4.288^{+0.018}_{-0.019}$ &575   \\
6.875&0.25  &$-5.190^{+0.049}_{-0.055}$  &175&5.9375&0.125 &$-4.384^{+0.021}_{-0.022}$ &470&5.9375&0.125  &$-4.337^{+0.018}_{-0.019}$ &592   \\
7.125&0.25  &$-5.463^{+0.036}_{-0.039}$  &153&6.0625&0.125 &$-4.460^{+0.021}_{-0.022}$ &443&6.0625&0.125  &$-4.395^{+0.018}_{-0.019}$ &547   \\
7.375&0.25  &$-5.441^{+0.036}_{-0.039}$  &142&6.1875&0.125 &$-4.702^{+0.023}_{-0.025}$ &336&6.1875&0.125  &$-4.556^{+0.021}_{-0.022}$ &462   \\
7.625&0.25  &$-5.511^{+0.039}_{-0.043}$  &131&6.3125&0.125 &$-4.874^{+0.026}_{-0.028}$ &282&6.3125&0.125  &$-4.737^{+0.022}_{-0.024}$ &399   \\
7.875&0.25  &$-5.80 ^{+0.12 }_{-0.17 }$  &93&6.4375&0.125  &$-4.985^{+0.028}_{-0.029}$ &262&6.4375&0.125  &$-4.704^{+0.041}_{-0.045}$ &356   \\
8.25&0.5    &$-6.241^{+0.084}_{-0.10 }$  &81&6.625&0.25    &$-5.088^{+0.024}_{-0.026}$ &329&6.625&0.25    &$-4.917^{+0.019}_{-0.020}$ &522   \\
8.75&0.5    &$-6.76 ^{+0.10 }_{-0.14 }$  &19&6.875&0.25    &$-5.047^{+0.025}_{-0.027}$ &307&6.875&0.25    &$-4.809^{+0.032}_{-0.035}$ &482   \\
& & & &                                   7.125&0.25       &$-5.080^{+0.023}_{-0.025}$ &346&7.125&0.25    &$-4.930^{+0.019}_{-0.020}$ &499   \\
& & & &                                   7.375&0.25       &$-4.940^{+0.030}_{-0.036}$ &398&7.375&0.25    &$-4.813^{+0.026}_{-0.028}$ &540   \\
& & & &                                   7.625&0.25       &$-5.268^{+0.023}_{-0.024}$ &390&7.625&0.25    &$-5.084^{+0.020}_{-0.021}$ &521   \\
& & & &                                   7.875&0.25       &$-5.358^{+0.049}_{-0.055}$ &261&7.875&0.25    &$-5.230^{+0.039}_{-0.043}$ &354   \\
& & & &                                   8.25&0.5         &$-5.833^{+0.029}_{-0.031}$ &246&8.25&0.5      &$-5.748^{+0.024}_{-0.026}$ &327   \\
& & & &                                   8.75&0.5         &$-6.180^{+0.064}_{-0.075}$ &57&8.75&0.5       &$-6.230^{+0.052}_{-0.059}$ &76    \\                            
\noalign{\smallskip}
\hline\noalign{\smallskip}
\end{tabular}
\end{table*}
\clearpage

\clearpage

\clearpage

\clearpage

\clearpage

\end{document}